\documentclass[11pt,twoside]{article}

\usepackage{asp2010}
\resetcounters
\begin{document}

\title{Sterrekundig Instituut Utrecht: The Last Years}
\author{Christoph U.~Keller
\affil{Leiden Observatory,
Leiden University,
P.O. Box 9513,
NL-2300 RA Leiden, 
The Netherlands}}

\begin{abstract}
  I describe the last years of the 370-year long life of the
  Sterrekundig Instituut Utrecht, which was the second-oldest
  university observatory in the world and was closed in early 2012
  after the Faculty of Science and the Board of Utrecht University
  decided, without providing qualitative or quantitative arguments, to
  remove astrophysics from its research and education portfolio.
\end{abstract}

\section{Good Times}

\subsection{Difficult Start for the New Faculty of Science}

In November 2007, Alfred Bliek became the new dean of the Science
Faculty of Utrecht University, a dean who turned out to be a strong
supporter of astronomy.  The merging of different departments into the
new Science Faculty a few years earlier had led to financial
difficulties because some departments brought very substantial,
structural deficits into the new faculty, thereby impacting other,
well-funded departments. Despite being financially sound and having
substantial reserves before the merge, the Physics \& Astronomy
department had gotten under financial pressure and was forced to lay
off some staff. No SIU staff was laid off. The reserves had been
blocked by the university board and directly or indirectly used to
cover the deficits of other departments. During these insecure times,
Norbert Langer, then director of SIU, won a prestigious Alexander von
Humboldt Professorship and departed at the end of 2008 for Bonn,
Germany.  As of 1.1.2009, the Faculty of Science had appointent me as
Scientific Director of SIU. Despite the word 'scientific' in the
title, most of the actual work had to do with administration,
bureaucracy and politics.

\subsection{A New Beginning}

In April 2009 the Sterrekundig Instituut Utrecht (SIU) embarked on an
ambitious plan, with full support from the dean. An additional staff
position was approved to reduce the heavy teaching load at the BSc
level. The position was filled by Maureen van den Berg in June
2010. The MSc program was enhanced with an instrumental track and more
space-related topics to attract students from technical universities,
which was reflected in its expanded name: Astrophysics and Space
Research. The plans for a new, common building with SRON, NOVA
(Nederlandse Onderzoekschool voor de Astronomie, the federation of
Dutch universities with astronomy departments) and the faculty’s
Scientific Instrumentation workshop started taking shape in the summer
of 2009. The new building would have allowed an effective sharing of
expensive machinery and cleanrooms between all partners. In the fall
of 2009, the Board of Utrecht University approved additional support
for collaborations with SRON (Netherlands Institute for Space
Research), providing SIU and IMAU (Institute for Marine and
Atmospheric research Utrecht) with 340,000 Euros per year for 5
years. In January 2010 the NOVA board expressed its plan to move its
Optical/IR group (10-15 FTE) to Utrecht around 2017/2018 to realize
the vision of Dutch astronomy to concentrate all its optical/infrared
instrumentation efforts in one place on a university campus. SIU and
NOVA were also ready to implement their plan to replace Norbert Langer
with a new full professor and two assistant professors, which was
presented to the faculty in October 2010. Most importantly, all these
plans were achievable with level funding from the university.

\subsection{Top Astronomy in The Netherlands}

Astronomy in the Netherlands is a top-scoring science. The 2010
evaluation of leading research schools reaffirmed this: NOVA and
Zernike (material sciences in Groningen) were labeled exemplary,
meaning above excellent. Indeed, Dutch astronomy is comparable in
performance to the top US institutes. This research excellence is
supported by excellence in education. SIU’s research was assessed as
nationally leading in its areas of expertise, internationally
competitive and making a significant contribution to the field. The
review committee recommended recruiting a world-leading researcher in
an exciting field of astrophysics and one or two additional faculty
hires at more junior levels, in agreement with SIU’s plans. In terms
of the faculty’s support of SIU’s ambition, the International Review
Board (IRB) noted: {\em “The IRB was very impressed with the clear vision
of the Utrecht University administration regarding the situation, and
hopes that the healthy collaborative atmosphere will lead to a
satisfactory resolution to the challenges facing the astrophysics
group.”}

\section{The Not So Good Times}

\subsection{Veerman Report}

In April 2010, the Veerman Report on Higher Education, a Report to the
Ministry for Education, Culture and Research prepared by a committee
of economists, managers and political scientists advised the Dutch
government to not have all universities focus on the same
high-priority research areas as this would cut the pie into too many
pieces. The report declared that a focus on top research was very
good, but to reach the top, the focus had to be coupled with choices
based on proven or wished-for strengths, collaboration, and not doing
everything anymore. In essence the Veerman report told the Dutch
universities to not compete with each other in high-priority research
areas. Astronomy, represented at five Dutch universities, could
therefore be viewed as one of the topics that too many universities
were involved in. This report obviously neglected the fact that
certain sciences, such as astronomy, are particularly attractive for
the natural and technical sciences, and that competition at the
national level is a crucial ingredient for top science. Furthermore,
NOVA already coordinated research to a significant extent: astronomy
in the Netherlands was geographically spread out, but nonetheless
quite coherent.

\subsection{Van der Eijk \& Klasen Report}

In June 2010, the report by Van der Eijk and Klasen, two chemists,
commissioned by the Board of Utrecht University on short notice,
provided external advise on how to solve financial and other perceived
problems in the faculty. The university board had obviously asked for
this report as it was not convinced that the plans of the Faculty of
Science would lead to a financially sound future. The report is full
of factual mistakes and ill advise. Indeed, the report was so poorly
written that many people, including me, simply ignored it. This may,
in hindsight, have been a mistake as it looks like the university
board appropriated many of the arguments in this report.

Without ever having talked to anybody in astronomy, the report
suggested national coordination ({\em “Zoek landelijke afstemming ten
  aanzien van de sterrenkunde, subatomaire fysica en theoretische
  fysica en vergroot bij deze onderdelen het omgevingsbewustzijn.”} In
English: Look for national coordination in the areas of astronomy,
particle physics and theoretical physics and increase their awareness
of other sciences), utterly unaware of the fact that NOVA plays
exactly this role, and does so exceptionally well. Furthermore, Van
der Eijk and Klasen were surprised by the small number of students
compared to the size of the staff ({\em “De adviescommissie is verrast
  over de relatief grote staf van het Sterrenkundig Instituut Utrecht
  in verhouding tot de geringe studentinstroom. De commissie adviseert
  verdergaande landelijke bundeling van het masteronderwijs en het
  onderzoek in de vorm van een sectorplan.”} In English: The advisory
committee is surprised by the relatively large staff of SIU in
relation to the small influx of students. The committee advices
further merging of national Master's education and research in the
form of a national plan), totally missing the point that astronomy
staff in the Netherlands are numerous because of their tremendous
research success, not because the Netherlands educates so many BSc and
MSc students in astronomy. Indeed, SIU most likely had more MSc
students per staff than any other Dutch university with an astronomy
MSc track. The committee had also missed the fact that NOVA had
already provided and implemented the plan they had advised.

\subsection{Science Faculty in Trouble}

On 14 June 2010, Alfred Bliek, dean of the Faculty of Science, stepped
down effective immediately because his vision and ambition for the
faculty was not shared by the university board. On 28 June 2010, the
retired psycho-pharmacist Jan van Ree was appointed as interim-dean and
stayed on until 1 January 2011, when the Utrecht chemist Gerrit van
Meer became the new dean. Based on the issues raised in the report by
van der Eijk and Klasen, van Ree appointed secret committees to advise
him on the different institutes.

The secret committee for SIU consisted of department and faculty
managers, lacking high-ranking scientists in physics or related
topics, all of them having serious conflicts of interest, and the
chair having shown a very negative and non-constructive attitude
towards SIU. In consultation with the permanent SIU staff, I gave a
presentation to the committee and provided information in writing. In
the end, the committee advised that astronomy should remain in
Utrecht, but that no expansion in whatever form would be possible
(even if fully funded from the outside), that astronomy should be
reduced in the number of research topics and that its MSc program
should be a joint effort with other universities and SRON. On top of
that, the committee had the audacity to advise that NOVA should
financially support astronomy in Utrecht when NOVA was already
providing a substantial amount of funding. This last suggestion
clearly showed that the committee had little or no understanding about
NOVA, or the organization of Dutch astronomy in general. Finally,
committee abused the NOVA institute rankings, obviously failing to
grasp that the ranking is given on an international scale, and not
(just) a national scale.

SIU objected with all means and sent a clearly-worded letter to van
Ree, who actually listened to the arguments. By December 2010, an
agreement (convenant) was signed by the university board and interim
dean van Ree as well as dean-to-be van Meer. This agreement provided
plans to restructure the faculty to reach a financially neutral result
by 2015. The efforts against the recommendations of the secret
committee had worked; regarding astronomy, the agreement contained the
following sentences: {\em “Astronomy will retain its current position
  and receive a stronger profile”} and {\em “Meteorology (IMAU) and
  Astronomy will merge in a Climate and Space research
  institute”}. While the latter made little sense, it would not have
had much of an impact except for destroying a well-known brand name.

By March 2011, van Meer had asked for a ranking of research groups by
the respective department heads. Research groups were never used in
Utrecht for anything, and they were indeed not well defined at all. In
some cases it was a single person, in other cases it was a large part
of an institute. The assessment criteria were highly subjective and
non-uniform. Coupled with rampant conflicts of interest (department
heads were also group leaders), these rankings led to a lot of bad
blood. I do not know what the result for astronomy or for any other
subject was as the results remain a secret until today. When I voiced
my criticism to the dean, van Meer, he answered with the following
email: {\em “Yes, as a scientist I respect the wish of scientists to
  be evaluated according to a well-defined objective, quantitative
  procedure. In contrast, most of the decisions that I will have to
  take concerning which units of the faculty we will discontinue do
  not strictly depend on such a procedure.”} This already indicated
that decisions would be made based on political motives and that
excellence in research and education or any other quantitative or
qualitative arguments would be of little relevance.

In addition to the rankings, highly regarded scientists in the
different departments were asked to provide assessments of the
different institutes. In April 2011 these ambassadors provided their
advise. I only saw a summary, which was favorable for astronomy: {\em
  “Astrophysics bestaat uit Experimental Astrophysics en High Energy
  Astrophysics. Hierin is aandacht voor observatie, theoretisch
  sterrenonderzoek, hoge energie astrofysica en instrumentatie. De
  keuze voor instrumentatie is een veelbelovende trekker voor de hele
  groep. Voor een stevig fundament voor de toekomst moet er wel
  gekeken worden naar de optimale positionering van High Energy. De
  groep kent sterke samenwerkings-verbanden en is gekwalificeerd als
  landelijke toponderzoeksschool.  Binnen de groep werken jonge,
  veelbelovende wetenschappers.”} And in English: Astrophysics
consists of Experimental Astrophysics and High-Energy
Astrophysics. They carry out observations, theoretical astronomy,
high-energy astrophysics and instrumentation. The choice for
instrumentation is a much appreciated motivator for the whole
group. For a solid foundation and the future, one needs to look into
how to optimally position high energy (astrophysics). The group has
strong collaborative ties and is qualified as a national top research
school. Young, highly acclaimed scientists are working within the
group.

Instead of revealing the decisions of the faculty board as planned on 26 May
2011, van Meer let the staff know the following: {\em “Ik heb zojuist een
gesprek gehad met het College van Bestuur over mijn voorgenomen keuzen
voor het toekomstig profiel. Dit heeft geleid tot nieuwe
gezichtspunten die maken dat ik een aantal zaken zal moeten
heroverwegen. Verdere beraadslaging is nodig om tot een zorgvuldige
afweging te komen.”} and in English: I just had a discussion with the
university board regarding my planned choices for the future (faculty)
profile. This has led to several new aspects that I need to
reconsider. Further discussions are needed to come to a careful
balance. My guess is that this was the time when the university board
told him to ax astronomy. But I have no proof for this. Indeed, it is
difficult to reconstruct what happened in the first half of 2011
as the research group assessments as well as the full ambassador reports
have been kept secret.

\section{The End}

\subsection{The Bomb}

On 8 June 2011, I was informed by the dean that the faculty wanted to
stop astronomy except for some teaching at the BSc level. No
quantitative or qualitative arguments were made, except for some
excuses like ``not the best institute in the Netherlands'', ``not
enough students'', and ``NOVA continuation not clear''. A planned
continuation of NOVA for 2013-2018 was at that time indeed not yet
approved by parliament.  I was told that all of this would be
implemented by laying off all SIU staff and that the decision to close
SIU was not firm, but that good arguments would be needed to reverse
it.

I immediately informed all permanent scientific staff, NOVA, NWO
(Netherlands Organisation for Scientific Research) , SRON and other
partners. Interventions from the highest national and international
levels started flowing directly to the dean and the university board.

Utrecht University, then the university that legally represented NOVA,
had decided to single-handedly withdraw from the well-coordinated
system of astronomy research and education in the Netherlands without
any advance consultation of any of the other stakeholders. Indeed, the
legal representative of one of only two exemplary research schools in
the country had decided to withdraw from the corresponding research
topic without consulting any of the other partners in the research
school, an act unheard of in history.

\subsection{Strategies}

The next day, 9 June 2011, we started discussing different strategies
to react to the university's plans. We saw two options:
\begin{enumerate}
\item a publicity offensive to keep astronomy alive in Utrecht
\item work out a large one-time payment by Utrecht University and use
  the money to move the staff to other Dutch universities
\end{enumerate}

The first option would have probably implied a smaller staff as vacant
positions could not have been filled anymore. Furthermore, nothing would
prevent the university from closing astronomy a few years later. The
trust in the university was lost, and it would have been difficult to
keep and attract good staff.

The second option had a number of clear advantages. It would
\begin{itemize}
\item provide a stimulating research environment in another location within 6 months
\item minimize uncertainties for SIU staff
\item make Dutch astronomy be seen as a unity
\item keep the research potential at the current level in the Netherlands
\item maximize the chances for NOVA continuation, considering the upcoming
  discussions in the Dutch parliament
\end{itemize}

Based on the clear advantages of the second solution, the SIU staff
uniformly supported exploring the second option and waiting with a
public outcry.

\subsection{Closed Doors}

The dean had mentioned that good arguments were needed to reverse the
decision. We certainly did not wait with presenting these arguments
and had them reinforced by high-level contacts at the national as well
as international level. But the faculty was not willing to listen. The
dean's secretary sent the following message to me on 9 June 2011: {\em
  “De afspraak die we vandaag gepland hebben voor morgenochtend met
  Gerrit van Meer kan helaas niet door gaan. Ik heb dit al op uw
  voicemail ingesproken. Gerrit van Meer heeft gisteren al met
  Christoph besproken dat hij eerst overleg moet hebben met de rector
  alvorens verdere afspraken gemaakt kunnen worden.”}. In summary, the
secretary let me know that van Meer was not available and that he
could not talk to anybody before consulting the rector, one of the
three members of the university board. The last sentence was incorrect
in that I had not been informed previously about this, something that
was indeed corrected by the dean in his email from 10 June 2011 to me:
{\em ''The sentence should have said that discussions at the level of
  NOVA and SRON will have to involve the rector (misunderstanding by
  the secretary).''}

Instead of waiting for an audience, on 10 June 2011, without an
appointment, Ewine van Dishoeck (scientific director of NOVA),
Wilfried Boland (executive director of NOVA) and I walked into the
dean’s office at 08:30 in the morning and asked him to meet with us
right away, to which he agreed. We pointed out the close national
collaboration through NOVA and that Utrecht University had to talk to
NOVA as well, in particular as Utrecht University had responsibilities
and obligations as NOVA's legal representative. Furthermore, we emphasized
the impact of the decision on NOVA, its credibility for future
international collaborations and the impact on NOVA's chances of being
continued beyond 2013.

The rector also blocked any discussion. On 10 June 2011 he wrote:  
{\em “Ik heb vanmorgen laten weten dat een gesprek op dit moment niet
zinvol is. Pas als het zicht op een definitief besluit wat verhelderd
is, kan ik hier (vanuit mijn perspectief) zinnig over spreken. Utrecht
zal bestaande verplichtingen uiteraard honoreren en in de
besluitvorming betrekken. Daarbij hoort ook dat wij partners op de
campus zoals SRON zo tijdig mogelijk informeren. Hierbij teken ik wel
aan dat vanaf volgende week de besluitvorming via een steil traject
zou kunnen verlopen, en dat een uitnodiging voor een gesprek van mijn
kant (b.v. op woensdag) noodgedwongen pas op dinsdagmiddag laat bij
jullie kan arriveren. Ik hoop dat jullie daar begrip voor hebben.”}
basically saying that discussions at this time would make no sense and
that he would only be available for discussions once the decision had
been finalized. This clearly indicated that the university board had
no interest in hearing any arguments against the closure.

\subsection{Respectful Divorce}

On 15 June 2011, Frank Verbunt and I met with the dean, van Meer. I
cite the dean from the minutes of this meeting: {\em “Ik ben tot dit
  voorlopig oordeel gekomen langs de bestuurlijke weg , waarbij
  verschillende oordelen die over jullie groep zijn gegeven zijn
  meegewogen. Volgens het rapport Veerman moeten we keuzes maken en
  het feit dat sterrenkunde op 5 plaatsen in Nederland bestaat, heeft
  invloed gehad op dit voorgenomen besluit.”  Ik heb moeite met dit
  besluit en heb geen kwalitatieve argumenten.''}  In English: I have
come to this preliminary judgement through a managerial process where
various judgements made regarding your group have been
considered, too. According to the Veerman report, we have to make
choices and the fact that astronomy exists in 5 places in the
Netherlands, influenced this preliminary decision. I have difficulties
with this decision and have no qualitative arguments.

It had become crystal clear that this was a political decision, most
likely by the university board and not the dean himself. In my
opinion, the president of the board, Yvonne van Rooy, wanted to show
the government that she was willing to take action and implement the
advice of the Veerman report quickly.

I then offered the dean our plan for a respectful divorce where we
would negotiate the terms of staff leaving voluntarily, given
substantial concessions by the faculty. The dean agreed to trying this
approach and to keep the discussions private until further notice.

On the same day, a delegation from NOVA and SRON met with the rector,
Bert van der Zwaan.  Wilfried Boland, Ralph Wijers from the University
of Amsterdam, Roel Gathier from SRON and I learned that the university
board's decision was firm and that there would certainly be no
negotiations to keep astronomy and invest through appointing new staff
on present, vacant positions. NOVA expressed its willingness to
negotiate a respectful divorce to secure the strength of the national
collaboration and keep the expertise in the country. NOVA also would
secure the completion and quality of ongoing PhD theses. The meeting
ended with a verbal agreement to keep things quiet and have NOVA and
the faculty negotiate a respectful divorce.

\subsection{Going Public}
On 16 June 2011 the faculty made its new, preliminary profile
public. The part on astronomy was very short: {\em ''Astrophysics niet
  in het profiel. Vervolgscenario na landelijk overleg.  Astrofysica
  blijft onderdeel bacheloronderwijs.''} (In English: Astrophysics is
not part of the (new) profile. Further developments await national
consultation. Astrophysics remains part of the Bachelor education.)
Interestingly, only few groups and institutes were mentioned as not
being part of the profile, and the cuts in research areas at that time
were far from sufficient to cover the deficit. The physics and
astronomy department would also loose its solar-cell research.

In the afternoon of the same day, I held an SIU all-hands meeting to
explain the events over the last week and the adopted strategy of
attempting a respectful divorce. Therefore, no public actions, press
releases or media events were planned.

The intention to close the Astronomical Institute in Utrecht was
reported in several national newspapers, and quickly became a topic of
discussion at scientific conferences.  Offers of help came flooding in
from all over the world, and many staff members had the difficult task
of explaining their colleagues that direct, vocal action would be
harmful.

During the week after the public announcement, we started to
implement the first actions towards a respectful divorce. On 17 June
2011 I informed the Graduate School of Natural Sciences of Utrecht
University about measures regarding teaching:
\begin{itemize}
\item all Bachelor students would be informed to obtain their Master's
  degree somewhere else,
\item current Master students could finish their degree in Utrecht, and
\item invoke a special rule that would allow us to hand out credits
  and grades outside of the regular curriculum.
\end{itemize}

On 22 June 2011 SIU issued a 'Dear Colleague' letter to the community
to explain the strategy. On 23 June 2011 we organized a meeting with
about 50 students where the other 4 universities presented their
programs and told students that they are exempt from the normal
registration deadlines.

Two weeks after the fatal, public announcement, we had taken charge of
the situation, including setting the agenda for all upcoming meetings
with the faculty. On 1 July 2011, Bram Achterberg, Ewine van Dishoeck
and I met with the faculty. We presented 58 pages of reactions to the
announcement, mostly from the international astronomical community as
well as a schedule for transitioning SIU to the other NOVA institutes
by the end of the year. A transition team was appointed, managed by
Felix Bettonvil (paid by NOVA) and faculty personnel in finance and
human resources. The faculty personnel was needed to obtain all the
relevant data for planning the transition, but they were not given any
information from our side.

\section{Transition}
On 30 September 2011, after several, sometimes frustrating, negotiations, we
reached a basic agreement. Utrecht would
\begin{itemize}
\item pay out about 5 years of salaries for all permanent staff who
  would transfer
\item not obtain any further funding from NOVA
\item let all PhD students obtain their degree at another university
\item let all equipment, computers, furniture etc.\ transfer to other
  universities at no cost
\item transfer all grants including Utrecht University-funded projects
  and pay until the end of the project
\item transfer all intellectual property of running projects at no cost
\item end all teaching obligations of SIU staff by the summer 2012
\item transfer the DOT to a foundation
\end{itemize}

Between October and December 2011 NOVA negotiated with other
universities to find positions for all permanent scientific
staff. This was a very difficult time for all staff and also their
families as things were not settled until shortly before Christmas. In
the end Jacco Vink was hired by the University of Amsterdam on a new
position, I received an offer from Leiden, and Heino Falcke made 4
positions in Nijmegen possible for Bram Achterberg, Frank Verbunt,
Soeren Larsen, and Onno Pols. Maureen van der Berg was offered a
temporary position at the University of Amsterdam. Marion Wijburg,
Sake Hogeveen, and Alexander Voegler, who has unfortunately been sick
for several years, will remain in Utrecht where the faculty is looking
for new jobs for them.

\subsection{Moving}

On 13 December 2011, the first part of the institute moved to
Amsterdam and Leiden. On 31 January 2012 the rest left for
Nijmegen. Only the wastebaskets were left in the empty rooms. At the
time of writing these sentences, the rooms were still empty. During
the same timeframe, SRON decided that Amsterdam would provide a better
environment for them in the future.

Only on 14 March 2012 did all 5 university boards sign the formal
agreement to transfer SIU to other Dutch universities. The last
accounts are being settled now.

In the end, Dutch astronomy will loose up to eight staff positions as
many of the positions created by the other universities will not be
structural. On the other hand, NOVA 4 was funded with 5 MEuro per year
for 5 years.

\subsection{This Conference}

On 17 June 2011, Vanna Pugliese asked me to support the idea of a
conference to honor SIU. Instead of a funeral, Vanna had the idea to
celebrate the 370-year life of SIU. There was widespread support among
the SIU staff, and Bram Achterberg suggested the title “370 years of
Utrecht Astronomy: from beginning to end”. The conference was
announced to SIU staff shortly afterwards.

On 5 September 2011, the SOC was formed, co-chaired by Vanna Pugliese
and Henny Lamers. The LOC was led by Marion Wijburg, who had organized
many of our conferences. Funding for the conference was made by
Utrecht University (we spent every last dime of the 2011 SIU budget on
paying for the venue and the meals), SRON, NOVA, and the LKBF.

\subsection{Thanks}
First of all, I thank all my colleagues at SIU, and also
their partners and families, for the amazing courage, patience and
solidarity that they have shown during these trying times. Bram
Achterberg and Frank Verbunt were most helpful and supportive in
meetings with the faculty, in analyzing the situation and taking quick
actions, and in giving priority to finding permanent positions for the
young staff members over their own search for new positions.
The NOVA Directorate, Ewine van Dishoeck and Wilfried Boland, was most
helpful, supportive and instrumental in making this transition
possible. My colleagues on the NOVA Board, Paul Groot, Thijs van der
Hulst, Koen Kuijken, Reynier Peletier and Ralph Wijers worked hard on
making positions available. Heino Falcke deserves special appreciation
for convincing the university board in Nijmegen to give 4 SIU staff a
new home.  Rens Waters and Roel Gathier at SRON provided advise and
support in these difficult times.

I would also like to acknowledge the tremendous support SIU received
from the astronomical community in the Netherlands and all over the
world. Finally, I much appreciate how the local and national
press understood our need for keeping things quiet while negotiating
with Utrecht University.


\end{document}